# Exotic single-photon and enhanced deep-level emissions in hBN strain superlattice

Xiang Chen[1], Xinxin Yue[1], Lifu Zhang[1], Xiaodan Xu[2], Fang Liu[1], Min Feng[1], Zhenpeng Hu[1], Yuan Yan[1], Jacob Scheuer[3], Xuewen Fu[1,4*]

**Affiliations:**

[1]Ultrafast Electron Microscopy Laboratory, The MOE Key Laboratory of Weak-Light Nonlinear Photonics, School of Physics, Nankai University, Tianjin 300071, China

[2]Key Laboratory for Microstructural Material Physics of Hebei Province, School of Science, Yanshan University, Qinhuangdao 066004, China

[3]School of Electrical Engineering, Faculty of Engineering, Tel Aviv University, Ramat Aviv, Tel-Aviv 6997801, Israel

[4]School of Materials Science and Engineering, Smart Sensing Interdisciplinary Science Center, Nankai University, Tianjin 300350, China

[*]Corresponding authors:  xwfu@nankai.edu.cn

**Teaser:** Realizing periodic modulation of defect-related single-photon and deep-level emission in hBN strain superlattice.

# Abstract

The peculiar defect-related photon emission processes in 2D hexagonal boron nitride (hBN) have become a topic of intense research due to their potential applications in quantum information and sensing technologies. Recent efforts have focused on activating and modulating the defect energy levels in hBN by methods that can be integrated on a chip, and understanding the underlying physical mechanism. Here, we report on exotic single photon and enhanced deep-level emissions in 2D hBN strain superlattice, which is fabricated by transferring multilayer hBN onto hexagonal close-packed silica spheres on silica substrate. We realize effective activation of the single photon emissions (SPEs) in the multilayer hBN at the positions that are in contact with the apex of the $SiO_2$ spheres. At these points, the local tensile strain induced blue-shift of the SPE is found to be up to 12 nm. Furthermore, high spatial resolution cathodoluminescence measurments show remarkable strain-enhanced deep-level (DL) emissions in the multilayer hBN with the emission intensity distribution following the periodic hexagonal pattern of the strain superlattice. The maximum DL emission enhancement is up to 350% with a energy redshift of 6 nm. Our results provide a simple on-chip compatible method for activating and tuning the defect-related photon emissions in multilayer hBN, demonstrating the potential of hBN strain superlattice as a building block for future on-chip quantum nanophotonic devices.

# 1. Introduction

Due to the characteristics of versatile optoelectronic properties, easy integration and tunability, two-dimensional (2D) van der Walls materials have become ideal platforms for studying various exotic physical phenomena across multiple fields of nanoscale science and technology, such as exciton funnel effect [1-3], formation of dark excitons [4], electron valley-polarized transport [5], single photon emissions (SPEs) [6], etc. It has been recently demonstrated that 2D materials of transition metal dichalcogenides (TMDs) and hexagonal boron nitride (hBN) can act as a solid-state host of single photon quantum emitters with high brightness, high purity and favorable quantum efficiency [7]. Among the various 2D hosts of quantum emitters, hBN is the most commonly used, as it can host defect-based room temperature quantum emitters with ultra-high brightness, absolute photostability, and robustness to harsh chemical conditions [8]. Its bandgap is approximatelly 6 eV [9-10], and it exhibits large defect tolerance that essentially benefits the tunable SPEs[11], making it a highly promising candidate for single photon sources – a key building block for quantum photonic network and quantum information technologies. Particularly, by introducing interband defects, such as boron vacancies ($V_B$), nitrogen vacancies ($V_N$), and substitutional antisite defects like $V_N C_B$ etc. [12] into hBN, a successful realizion of SPEs spanning across ultraviolet (UV), visible (VS), and infrared (IR) spectral ranges [13-17] have been demonstrated. These defect-level structures exhibit a wide range of tunable photophysical properties for the SPEs, including emission brightness, wavelength, linewidth, stability and chirality [18-22]. In addition, deep-level (DL) UV emission is another important defect-related optical property of hBN, which was once mistaken as band edge emission due to its strong intensity at room temperature [8]. It has been widely demonstrated that the DL emission intensity can be tuned by controlling the twist angle of stacked hBN multilayers and the external voltage [23-24]. Furthermore, due to the high mechanical strength and flexibility, 2D hBN can be well integrated with cavities, photonic waveguides and plasmonic structures, making it suitable for applications in on-chip quantum nanophotonic devices. Therefore, developing an approach for activating and regulating the defect-levels in hBN for the realization of stable and high efficient SPEs in a controllable manner, which is compatible with on-chip integration process, is a fundamentally important task which is attractive for applications in quantum information science.

Several different methods have been explored to activate and regulate the defect-levels in hBN for realizing SPEs. These include doping by CVD growth [25], ion radiation [26], electron beam radiation[14, 16], twisted stacking [23-24], applying external electric [18] and magnetic fields [22] etc. However, the SPE location distributions in the hBN, created using these methods are generally random, and controllable creation of efficient SPEs is still challenging. Elastic strain offers a novel and powerful tool to tailor the physical properties of various 2D materials due to their high mechanical strength and flexibility[27]. For example, site-controllable SPEs in $WSe_2$ have been obtained by elastic strain engineering [28]. It has also been demonstrated that local elastic strain can effectively activate room temperature SPEs in hBN [17], and that strain regulation allows for spectral tunability over 65 meV [19]. Furthermore, implementation of controllable and device-compatible strain patterns for activation of SPEs is even more important because it facilitates the realization of on-chip quantum nanophotonic devices [29-31]. Recently, such concept has been successfully employed to fabricate strain patterned molybdenum disulphide ($MoS_2$) devices to enhance the exciton emissions [32], which, in principle, would be also applicable for creating hBN strain superlattice for patterned SPEs due to its high mechanical strength and flexibility [11]. Such artificial hBN strain superlattice is achievable by traditional planar process, which can be well integrated with optical cavities and photonic waveguides for large-scale and device applications [31, 33]. However, systematic investigation on artificial hBN strain superlattice assembly is still lacking, and the microscopic mechanism of the strain effect on the defect-levels in hBN is still elusive.

Here, we report a systematic study of defect-related photon emissions in hBN artificial strain superlattice ensembles that are formed by transferring multilayer hBN onto hexagonal close-packed silica spheres on silica substrate. Using combined photoluminescence (PL) and cathodoluminescence (CL) measurements, local strain dependent periodic tunable optical properties of the defect-levels in the hBN artificial strain superlattice ensembles are revealed. First, under the excitation of a 532 nm laser, we observe bright and stable room temperature SPEs in the strain superlattice, and the emission peaks of the zero phonon line (ZPL) and the photon sideband (PSB) are effectively modulated by the periodic local strain with a spectral blueshift of 12 nm and 13 nm, respectively. Second, through high spatial resolution CL measurements, we find that the periodic local strain can dramatically enhance the intensity of the DL emission in the hBN , where the maximum enhancement is up to 350%, accompanied with a spectra redshift of 6 nm. Third, comparing the PL and CL results on both the SPEs and DL emissions in the hBN strain superlattice and in combination with first-principles calculations, we reveal that the observed SPEs and DL emissions stem from the same defect classes, and provide a physical picture for the periodic local strain regulation on the defect related photon emissions. Our work realize periodic strain-modulated defect-level emissions in hBN via a simple artificial superlattice, which would be a potential platform for designing scalable light quantum sources compatible with the on-chip integration process.

## 2. Results and Discussion

**Preparation of hBN strain superlattice**

For the preparation of hBN strain superlattice sample, multilayer hBN sheets (from several to tens of nanometers in thickness) were mechanically exfoliated from a bulk hBN crystal (Taizhou SUNANO New Energy Co., Ltd.) using standard scotch tape techniques and deposited on a $SiO_2$ (80 nm) /Si substrate. Then the multilayer hBN sheets were transferred to a pre-patterned silica microspheres (MSs) substrate (see **Methods**) by a wetting transfer process. The detailed transfer steps are shown in **Fig. 1(a)**. First, 200 nm polymethyl methacrylate (PMMA) film is spin-coated onto the hBN sheets on the $SiO_2$/Si substrate and the substrate beneath is then removed by NaOH (5mol/L) solution etching at 85℃. Second, spin coat monodispersed silica MSs (~1 μm in diameter) solution onto a fresh $SiO_2$ (80 nm thick) /Si substrate to form monolayer hexagonal close-packed silica MSs arrays. Then, this pre-patterned substrate is used for picking up the hBN sheets floating in the deionized water. After several cycles of deionized water cleaning, the hybrid structures are immersed in acetone to remove the PMMA and are dried in vacuum, yielding multilayer hBN strain superlattice structures. Finally, the hybrid structures are annealed at 200℃ for 2h. **Fig. 1(b)** shows the typical optical (top) and scanning electron microscopy (SEM) (bottom) images of a representative hBN strain superlattice structure. Silica MSs were chosen due to their simple preparation, clean surface, insulating properties, well-controlled spherical shape and compatibility with the planar process [34]. Note, that large-scale monolayer of hexagonal close-packed MSs are easily accessible without requiring lithography or ion beam figuring. After transferring the multilayer hBN sheets onto the top of the prepared MS substrate, both strained (on the MSs domains) and unstrained regions (on the flat silica substrate) are obtained in the same hBN sheet.

**Fig. 1(c)** shows a side view of the schematic diagram of the hBN strain superlattice. Due to the strong capillary force between the hBN sheet and the substrate during the drying process, the flexible hBN sheet will be firmly confined to the spherical surface of the silica MSs with an unavoidable geometric conflict. Consequently, the hBN sheet acquires substantial mechanical stresses. Specifically, the hBN sheet withstands remarkable biaxial tensile strains at the contact areas between the hBN sheet and the silica MSs, resulting in the periodic tensile strain distribution in the hBN sheet, namely, the 2D hexagonal hBN strain superlattice (**Fig. 1b**). Note that the hBN sheet mainly adheres to the apex of the silica spheres and is suspended between the spaces of the adjacent silica MSs. Besides the large-scale array regions, we could also obtain single silica sphere and multi-sphere (>1) regions (**Fig. S1**). Such different configurations of the silica MS assembly enable comprehensive study on the complex strain effect on the defect-levels in the hBN.

**Strain activation of SPEs in hBN strain superlattice**

To study the strain activation of SPEs in hBN strain superlattice, we used scanning confocal microscopy with a 532 nm laser excitation to investigate its PL and Raman spectra (see **Methods**), in which the diameter of the focused laser spot is less than 1 μm. **Fig. 2(a)** shows the morphology of the studied sample, where the green-colored spot 1 and the blue-colored spot 2 indicate representative unstrained region of the hBN on the silica substrate and the strained region in the hBN strain superlattice, respectively. To evaluate the strain in the hBN superlattice, we measured the Raman spectra at both the strained (spot 2) and unstrained regions (spot 1). As shown in **Fig 2(b)**, the frequency of $E_{2g}$ mode at the unstrained region is 1364.9 cm$^{-1}$, in line with a previous report [35]. In contrast, the frequency of the $E_{2g}$ mode redshifts to 1359.2 cm$^{-1}$ at the strained region, indicating the existence of a substantial biaxial tensile strain in the hBN strain superlattice. From the frequency redshift of the $E_{2g}$ phonon mode, the magnitude of the tensile strain in the hBN strain superlattice is determined to be ~ 0.15% [36-37].

As pointed out in previous studies, hBN is expected to host optically active defects that have ground and excited states within the gap [38], which are usually inactivated unless external stimulation is applied [13-14]. To identify the state of such defects, we conducted PL measurements with sub-bandgap excitation (wavelength of 532 nm, power of 1.07 mW) at both the strained (spot 2) and unstrained (spot 1) regions (**Fig. 2(c)**). Intriguingly, the strained region displays intensive photon emissions in the spectrum range from 560 nm to 780 nm, while no observable photon emission were observed in the unstrained region. The observed main emission peak at 624.2 nm (1.99 eV) in the strained region matches well with that reported ZPLs of the SPEs in hBN. The broader peak at 678.2 nm, ~158.2 meV away from the ZPL, is recognized as a phonon sideband (PSB) related to the in-plane optical phonons of hBN [39]. Note that the sharp emission at 571 nm in the PL spectra stems from the Raman scattering related to the $E_{2g}$ phonon mode [35, 40]. To eliminate the possibility of insufficient excitation efficiency, PL spectra under different excitation power levels were collected from both the strained and unstrained regions (**Fig. S2**). The results show that regardless the excitation power, the defects do not exhibit any activity in the unstrained region, while the photon emission intensity in the strained region increases linearly with the excitation power, thus confirming that the defect levels are effectively activated in the hBN strain superlattice. **Fig 2(d)** shows the confocal PL intensity map of the whole hBN flake using the wavelength range of the SPEs (from 600 nm to 700 nm), which follows well the profile of the hBN strain superlattice area, again demonstrating the efficient strain activation of the SPEs in the hBN strain superlattice. In addition to the strain superlattice area, the interfacial area that connecting the suspended hBN and the hBN absorbed on the silica substrate also exhibits SPE, indicating substantial local tensile strain along this interface. Note that the measured SPE intensity distribution is not uniform due to the periodic strain distribution in the hBN strain superlattice. Because of the finite spatial resolution of our experimental setup, the SPE intensity map cannot show clearly the hexagonal strain pattern in the hBN strain superlattice (**Fig. S3**). Therefore, we clearly demonstrate the activation of the defect-level related SPEs in the hBN strain superlattice.

**Modulation of SPEs in strained hBN over a single silica MS**

To further reveal the relation between the SPE optical properties and the local tensile strain in the hBN strain superlattice, we systematically investigated the Raman and PL spectra of a strained hBN flake over a single silica MS by linearly scanning the laser spot across the MS. The left panel of **Fig. 3a** displays schematically the cross-sectional diagram of a single silica MS covered by a thin hBN flake. Due to the symmetrical structure of the ensemble, the spatial distribution of the local strain is also symmetrical around the MS. Along the arrow from the left edge through the center of the MS to its right edge, the strain in the hBN flake increases gradually, reaching a maximum at the apex of the MS and then begins to decrease gradually [41]. The right panel of **Fig. 3a** shows a typical optical image of the strained hBN flake over a single silica MS, in which the arrow indicates the scanning direction of the laser spot. **Fig. 3b** presents the corresponding line-scan Raman result. As guided by the blue dashed arrow, the frequency of the $E_{2g}$ mode shows a continuous redshift until arriving at the apex of the MS and then gradually blueshifts back, indicating the maximum tensile strain in the hBN locates at the apex of the MS. **Fig. 3c** presents the corresponding line-scan PL result. As indicated by the pink and green dashed arrows, both the ZPL and the PSB emission peaks exhibit continuous blueshift first and then gradually redshift back, with the maximum blueshift at the apex of the MS. Therefore, in addition to the activation of SPEs in hBN, the emission energy of the SPEs can also be effectively turned by the biaxial tensile strain.

To quantify the local tensile strain modulation effect, **Fig. 3(d)-(e)** present Gaussian-fitting retrieved peak positions of $E_{2g}$, ZPL and PSB as a function of the laser scanning positions across the MS. All of the above show a nearly linear dependence at the tensile strain region. The maximal redshift of $E_{2g}$ is 2.5 cm$^{-1}$, and the maximal wavelength blueshifts of ZPL and PSB are 12 nm and 13 nm, respectively. Using the strain potential of hBN in previous work [35], the maximum tensile strain in the strained hBN at the apex of the MS is estimated to be ~0.06%. As the spot size of the laser is ~1 μm and the step size of the linescan is 150 nm in our measurements,, each measured point represents the average optical information from an excited area of ~1 μm$^2$. Since the diameter of the MS is only 1 μm, such limited spatial resolution of the line-scan Raman may lead to an underestimation of the real maximum strain in the hBN flake. Nevertheless, our results clearly show that a tiny biaxial strain can effectively activate and modulate the SPEs in hBN, which possess potential in designing hBN strain superlattice with pattern controllable and energy tunable single photon emitters that compatible with the on-chip integration process.

**Modulation of DL emission in strained hBN flake over a single silica MS**

In addition to the SPEs, DL UV emission is another important, defect-related, optical property of the hBN due to its potential application in UV light source for photonic devices [8], which is also sensitive to external stimuli such as electric field and stacking twist angle [23-24]. In the following part, we turn to the investigation on the DL emission in the hBN strain superlattice by using high spatial resolution CL. First, we studied the strain modulation on the DL emission by CL measurements on a strained multilayer hBN flake over a single silica MS. The typical scanning electron microscope (SEM) image of the sample is shown in **Fig. 4(a)**. As reported previously, the intensity of the second electron signal of the SEM image is sensitive to the orientation of the incident beam with respect to the lattice surface, thus the local contrast difference in the SEM image can qualitatively reflect the presence of lattice deformation [42]. Due to existence of substantial lattice deformation and strain of the hBN flake in the wrinkles and region over the MS, they shown much brighter contrast than that of the suspended regions and on the flat silica substrate in the SEM image (**Fig. 4(a)**). **Fig 4(b)** presents representative CL spectra of the unstrained hBN (region 1 on the flat substrate), less strained hBN (suspended region marked by 3), and most strained hBN (region at the apex of the MS marked by 2), where the broad CL band from 300 nm to 450 nm arises from the DL emission of hBN. In **Fig. 4(b)**, the CL spectra of the silica substrate and the silica MS collected under the same measurement condition are also shown to exclude any possible interference from the background. The CL intensity of the silica substrate is nearly zero and negligible. Although the silica MS exhibits some CL signals at ~400 nm and ~650 nm, their intensities are at least one order of magnitude lower than that of the hBN, and there is only a tiny overlap in the emission range with the hBN, thus the observed CL in the UV band mainly comes from the DL emission of the multilayer hBN. Intriguingly, the DL UV emission of the multilayer hBN shows a significant intensity enhancement with increasing the tensile strain, with the maximum enhancement at the most strained area at the apex of the SM. **Fig 4(c)** shows the DL emission intensity map of the area indicated by the blue-dashed rectangular in **Fig 4(a)**. A strong localized DL emission is observed in the MS region, and the intensity distribution follows very well the geometry of the MS with the maximum CL intensity at the apex. The origin of the DL UV emission in hBN has been controversial for a long time. Recently, the carbon-related defects (color centers) have become the widely recognized explanation [43]. Therefore, the observed enhancement of the DL UV emission in the multilayer hBN flake indicates that the elastic tensile strain can substantially modulate the optical transition and radiation process of the carbon-related color centers in hBN.

To quantitatively analyze the strain enhancement of the DL emission in the hBN, we have collected a set of line-scan CL spectra on the strained multilayer hBN by scanning the focused electron spot ,step by step, across the silica MS, as indicated by the orange arrow in **Fig. 4(a)**. The raw line-scan CL spectra are presented in **Fig. 4(d)**, in which the orange arrow denotes the line-scan direction of the electron spot. Clearly, with the excitation electron moving towards the apex of the MS, the DL UV emission intensity of the hBN gradually increases and reaches the maximum intensity at the apex of the MS, where the hBN has the maximum biaxial tensile strain. In parallel, the peak position of the DL UV emission shows a gradual redshift. With the electron excitation further moving away from the MS, the DL UV emission intensity and wavelength gradually return back to their initial values. Since the tensile strain in the hBN is largest at the apex of the MS and gradually decreases towards the perimeters of the MS, thereby, the CL intensity of the DL emission follows the same trend. The quantitative spatial evolution of the DL emission intensity and peak position across the MS can be seen more clearly by the extracted peak intensities and positions from the line-scan CL data, as plotted in the top and bottom panels of **Fig. 4(e)** as a function of the electron spot scanning position. The maximum intensity enhancement and wavelength redshift of the DL emission are estimated to be more than 350% and ~ 6 nm at the most strained region. Note that the observed redshift of the DL emission is contrary to the observed blueshift of the SPEs in the strained hBN flake (see **Fig. 3**), indicating the different mechanisms of the strain effects on the SPEs and DL emission, which will be discussed later. To rule out the possible interference of sample diversity, several different strained hBN flakes have been studied and all of them show the similar result (see **Fig S4**), demonstrating the high reproducibility of our strain-loading method and the strain enhancement of the DL UV emission in hBN is an intrinsic property.

There are several possible factors may enhance the CL intensity for the material under elctron excitation, including varying local emission geometry [44-45], varying substrate optical constant [46], strain-induced carrier/exciton funneling [1, 3, 32], and strain-enhanced radiative recombination [47]. First, since the excitation region of the electron beam is only about tens of nanometers (**Fig S5**), much smaller than the diameter (1 μm) of the silica MS, the geometry of the hBN flake at the apex region of the silica MS can be regarded as nearly flat in the CL test and thus the local geometry effect can be neglected. Second, the substrates of the strain-free region (silica film) and strained region (silica MS) are composed of $SiO_2$, the impact of varying substrate optical constant can be excluded. Third, according to the Raman result the biaxial tensile strain is relatively small (~0.1%) even at the apex of the MS, such small built-in field is insufficiently to drive the carriers towards the apex spanning a range of hundreds of nanometers. Therefore, we speculate the observed > 350% CL intensity enhancement is mainly due to the strain-enhanced radiative recombination process, where the non-unifrom biaxial tensile strain could substantially modulate the energy bands and transition rule [48-49]. Such similar phenomena have been observed in other traditional semiconductors [47]. For the hBN, the orbital wave function of the defect level is sensitive to external stimuli including the elastic strain. Applying biaxial tensile strain will change the relative energy position of the defect level to the valance and conduction bands, resulting in modification of the electron occupancy in the corresponding energy level and thus the enhanced DL photon emission [20, 50-51].

**DL emission intensity mapping of different artificial hBN strain superlattice assemblies**

We further studied the DL emission intensity mapping of different hBN strain superlattice assemblies, in which the periodic strained pattern of the hBN strain superlattice assemblies were revealed by CL mapping measurements with high spatial resolution. **Fig. 5(a)** shows the hexagonal close-packed silica MSs arrays uncovered (top) and covered (bottom) by multilayer hBN flake, where the boundary is indicated by the blue dashed line. **Fig .5(b)** shows a set of line-scan CL spectra collected along the blue arrow indicated in **Fig. 5(a)**, which show periodic variation following the closed-packed silica MSs in both the CL intensity and wavelength. **Fig. 5(c)** shows a typical DL UV emission intensity map of the multilayer hBN over the close-packed silica MSs by monochromatic CL measurement, in which the periodic DL emission intensity distribution precisely copies the hexagonal morphology of the close-packed silica MSs with the maximum emission intensity at the apex of each MS, directly evidencing the enhancement of DL emission by the two dimensional strain pattern in the multilayer hBN endowed by the MS substrate. In addition, we further implemented polychromatic CL measurements on different types of hBN strain superlattice assemblies, which has the advantages for fast and large area CL mapping test. **Fig. 5(d)-(f)** present the SEM images of the different hBN strain superlattice assemblies, and their corresponding polychromatic CL mapping images are shown in **Fig. 5(g)-(i)**. Apparently, the CL emission enhanced at the strained region over the silica MSs and the spatial distribution of the CL intensity is consistent with the morphology of SEM image. The most strained areas of the multilayer hBN that contacting with the top surface of the MSs exhibit much higher CL brightness than that of the suspended areas with less strain, while the strain-free areas at flat silica substrate nearly have on observable CL signals, demonstrating the powerful capability of the silica MS based artificial hBN strain superlattice for local and patterned enhancement of DL emissions in multilayer hBN, which has the potential to construct large-scale functional photonic devices with spatially controllable and enhanced DL emissions in the UV range.

To understand the biaxial tensile strain regulations on both the SPEs and DL emissions in the hBN strain superlattice observed by the PL and CL measurements, we examed the variation of the atomic defect structures and the related energy-levels of the SPEs and DL emissions under strain. The atomic structure and the related energy-level of the SPEs in hBN are complicated and still elusive, but recent studies suggest carbon as a key component for the SPEs [25, 52]. Meanwhile, the DL UV emission has also been widely recognized as steming from the carbon impurities, which are named as color centers [24]. In our experiments, carbon impurities are inherent in the hBN crystal (introduced during the synthesis), thus we can observe the slight DL emission in the unstrained region. While the SPE is naturely inactive [29]. Under the biaxial tensile strain the SPE in the multilayer hBN is efficiently activated, meanwhile the DL emission is remarkably enhanced (**Fig. 2** and **Fig. 4**). Moreover, the SPE and the DL emission show opposite energy shift, i.e. blueshift (38.6 meV) for the SPE while redshift (49.6 meV) for the DL emission, and their opposite photon energy shifts are roughly equal. In addition, the energy sum of the DL emission (3.54 eV) and SPE (2.0 eV) is approximately equal to the band gap of the hBN. Therefore, it is reasonable to suspect that the observed SPE and DL emission would stem from the same class of defect levels, but through their different optical transition processes.

Based on the above analysis, the possible schematic transition diagram for the SPE and DL emission is depicted in **Fig. S6.** With a sub-bandgap laser excitation, the electrons are directly excited from valence band to the defect levels and the subsequent recombination of the excited carrier result in the SPEs. While under a high energy electron beam excitation, the valence band electrons will be firstly excited to the bottom of the conduction band and then transit to the defect levels through a radiative recombination process that corresponding the DL emission in the UV range. Under the tensile strain modulation, the defect energy levels would move closer to the bottom of the conduction band, resulting in the opposite energy shift of the SPE and DL emission. To prove this hypothesis, we performed first-principle calculations based on density function theory (DFT) to model the change of the defect energy levels induced by the bilaxial tensile strain in bilayer hBN (see **Methods**). According to the previous studys [52-53], $V_BC_N$ defect is selected as a representative structure in our calculations. As shown in **Fig. S7**, under 1% biaxial tensile strain the energy levels of the $V_BC_N$ defect are notabley modified, where the mid-band defect levels move closer to the conduction band while the position of other energy levels change barely. Therefore, the photon emission originating from the transition between the ground state or valence band and the mid-band defect levels, namely, the SPEs will show a blueshift, while the DL emission stemming from the transition between the conduction band and the mid-band defect levels will exhibit a redshift, verifying the physical picture described in **Fig. S6**.

## 3. Conclusion

In conclusion, through systematic PL and CL measurements of silica MS based artificial hBN strain superlattice, we observed the exotically activated SPE and enhanced DL emission in the hBN. The photon energies are modulated by the local biaxial tensile strain in an inverse manner. Based on the experimental results, we provided a compelling evidence for the physical picture that the SPEs and DL emissions originate from the same class of defect levels with different transition processes, which are sensitive the tensile strain. Furthermore, on-chip device compatible strain patterns have been realized in multilayer hBN by a simple method without requing a lithographic process, in which the strain superlattice is endowed with the periodicity of the pre-patterned silica MSs. This demonstrated hBN strain superlattice is generally applicable to other 2D materials with spatially textured optical properties. Our result not only sheds light on the strain modulation on the photophysical properties of the color centers in multilayer hBN, but also provide a promising paradigm for future design of multifunctional quantum photonic devices.

## 4. Methods

*Preparation of monolayer SiO$_2$ sphere assemblies:* Monodisperesed silica micropheres (MS) solution were purchased from HWRK CHEM (M814155-5ml) with the MS concentration of 2.5%. 80 nm SiO$_2$ / Si substrates were cleaned via sonication in acetone and IPA, and oxygen plasma, following which silica MSs solution were spin-coated at 500 rpm for 6 s and 2000 rpm for 30 s. Thereafter, the samples were baked at 150 °C for 10 minutes on a hotplate to remove the residual water. Silica MSs will assemble together, forming separated monolayer regions with close-packing order, range from a few microns to tens of microns.

*Raman and photoluminescence characterizations:* The Raman and PL spectroscopy measurements were carried out using a HORIBA LabRAM ODYSSEY system with a 532 nm laser for excitation. The spot size of the focused laser is ∼1 μm. The signal was collected by a 100 X objective lens with a numerical aperture of 0.9. 1800 lines/mm and 100 lines/mm gratings were used for Raman and PL measurements, respectively.

*CL measurements:* CL spectra and mapping were measured using a HORIBA H-CLUE system integrated to a Thermofisher Quattro S SEM. A focused electron beam with beam current of ∼46 pA (electron energy of 10 keV) scanned the samples while recording the CL emission spectrum synchronously to produce hyperspectral images. The emitted light was collected by a parabolic mirror and sent to a UV-vis spectrometer (Horiba iHR320) equipped with a 150 lines/mm grating using a slit width of 200 μm, then the CL signal was directed to a thermoelectrically cooled silicon charge coupled device (CCD) array (Synapse Plus). The SEM images were obtained with a standard Everhart–Thornley detector for collecting secondary electrons.

*DFT calculations:* All the calculations were performed based on density functional calculations as implemented in the Vienna Ab initio Simulation Package (VASP) [54-57]. Plane-wave basis sets with energy cutoff of 520 eV were used and Perdew-Burke-Ernzerhof (PBE) functional was adopted in the calculations [58]. The van der Waals interactions is included using the semi-empirical DFT-D3 type of dispersion energy correction [59]. All the structures were relaxed until the energy was lower than $10^{-5}$ eV, and the force was less than 0.02 eV/Å on each ion. Owing to the underestimation of the energy gap by PBE functional, the band structures are calculated with the Heyd-Scuseria-Ernzerhof hybrid functionals (HSE06) [60].

**SUPPORTING INFORMATION**

Figures S1 to S8

**Data availability**

The data that support the findings of this study are available from the corresponding authors upon reasonable request.

**Acknowledgments:**

This work was supported by the National Natural Science Foundation of China (NSFC) at grant No. 11974191 and 2217830, the National Key Research and Development Program of China at grant No. 2020YFA0309300, the Natural Science Foundation of Tianjin at grant No. 20JCZDJC00560 and 20JCJQJC00210, the 111 Project at grant No. B07013, and the "Fundamental Research Funds for the Central Universities", Nankai University (grant No. 91923139, 63213040, C029211101, C02922101, and ZB22000104).

**Author contributions**

X. F. conceived the research project. X. C. and X. Y. did the experimental measurements. X. C. proceeded data analysis. L. Z. and Z. H. performed the density functional theory calculations. X. F. and X. C. wrote the manuscript. All the authors contributed to the discussion and revision of the manuscript.

**Competing financial interests**

The authors declare no competing financial interests.

**Figure legends**

**Figure 1. Fabrication and micrograph characterization of multilayer hBN strain superlattice sample.** (a) Schematic diagram of the detailed fabrication process for hBN strain superlattice by pre-pattern wetting transfer technology. (b) Typical optical (top panel) and SEM images (bottom panel) of the prepared hBN strain superlattice sample. (c) Schematic diagram of the hBN periodical strain superlattice. The diameter of the silica MSs is 1 um. Due to the van der Waals force between the hBN sheet and the MS substrate, biaxial tensile strain is applied to the hBN sheet.

**Figure 2. Raman and PL spectroscopy measurements of the hBN strain superlattice.** (a) Optical image of a hBN strain superlattice sample. The white dotted line is the boundary of the hBN sheet. Points 1 and 2 represent the unstrained and strained regions for measurements, respectively. (b) Raman spectra of the two representative areas indicated by points 1 and 2 in (a). The peaks were fitted with a Gaussian function. (c) Porrespoonidng PL spectra of the two representative areas. For the PL spectrum at the strained area, two peaks were determined by the Gaussian fitting and they corresponds to the ZPL and PSB peaks centered at 624.2 and 678.2 nm, respectively. (d) PL intensity map with the SPE emission (600 $< \lambda <$ 700 nm) from the same area shown in the optical image.

**Figure 3. Line-scanning Raman and PL measurements of a strained hBN flake over a single MS.** (a) Schematic diagram of the hBN flake over a single MS. Right inset: a typical OM image of the sample. (b) Line-scanning Raman spectra of the hBN collected at the strain region along the black arrow indicated in (a). The blue arrow showed the evolution of $E_{2g}$ peak position. (c) Corresponding line-scanning PL spectra of the hBN collected at the strain region along the black arrow indicated in (a). The pink and the green arrows show the evolution of ZPL and PSB peak positions, respectively. (d)-(f) The position (strian) dependent peak positions of $E_{2g}$, ZPL, and PSB extracted from the (b) and (c).

**Figure 4. High spatial resolution CL measurements of strained hBN flake over a single MS.** (a) SEM image of the strained hBN flake over a single MS for analysis. (b) The typical CL spectra collected at position 1 (unstrained region), position 2 (most strained region), and position 3 (less strained region) of the strained hBN flake indicated in (a), in which the purple solid and dashed curves are the CL spectra collected at other positions of the substrate for reference. (c) Line-scanning CL spectra of the strained hBN flake across the below MS along by the orange arrow in (a). (d) CL intensity map with the DL emission ($300 < \lambda < 400$ nm) in the strained hNB flake over a single MS, where the CL internsity map follows well the morphology of the MS. (e) Position (strian ) dependent total DL emission intensity obtained by integrating the CL spectrum from 300-400 nm (top panel) and dominant peak position retrieved by Gaussian fitting (bottom panel).

**Figure 5. High spatial resolution CL and polychromatic CL imaging measurements of different hBN strain superlattice assemblies.** (a) SEM image of a periodic hBN strain superlattice sample measured at 10 kV. The light blue dotted line is the boundary between the $SiO_2$ sphere array covered and uncovered by the hBN sheet. Dashed hexagon revealing the local periodic pattern. (b) Line-scanning CL spectra along the blue arrow indicated in (a), where the color from red to blue represents the CL intensity from strong to weak. (c) CL intensity mapping by the DL emission at the period hBN strain superlattice. (d)-(f) SEM images of three types of hBN strain superlattice assemblies. (g)-(i) The corresponding polychromatic CL images of the samples in (d)-(f). The contrast between the strained region and unstrained region highlights the remarkable enhancement effect of the local strain on the DL emission of hBN.

**Figure 1**

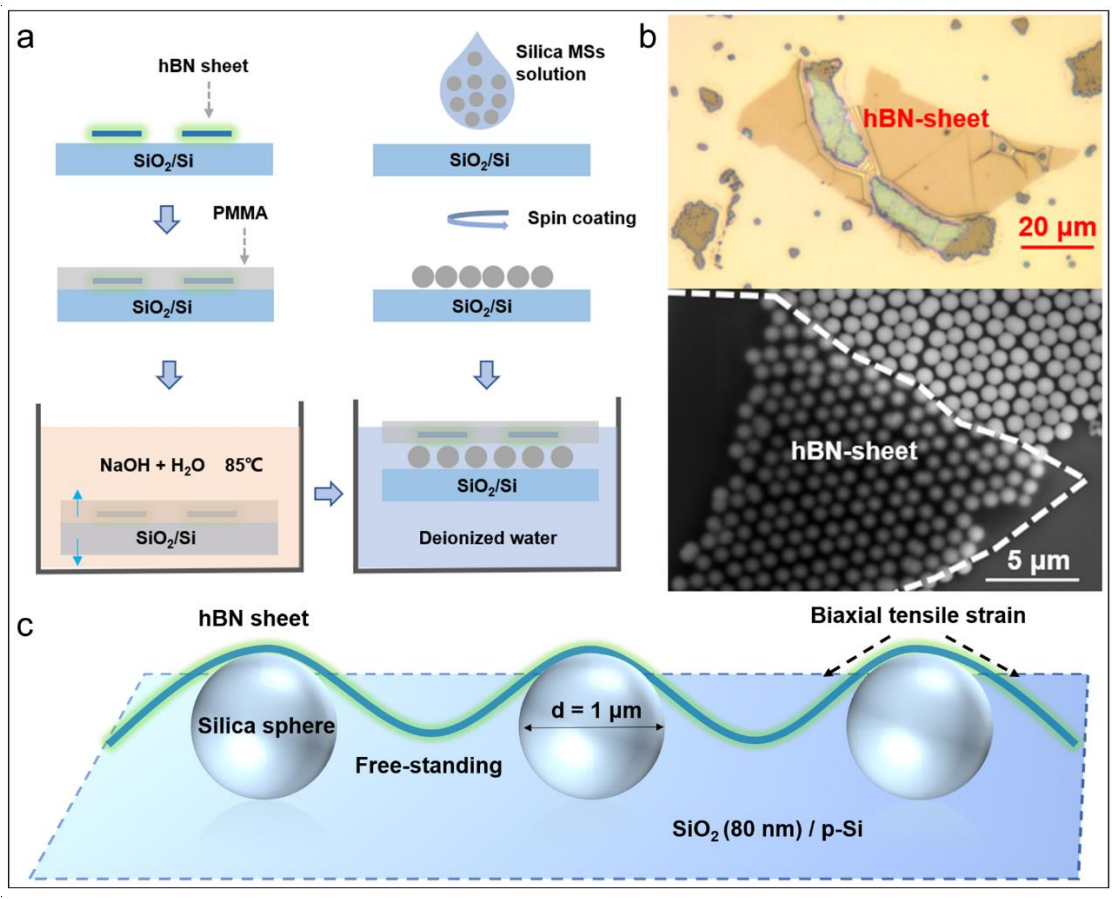

**Figure 2**

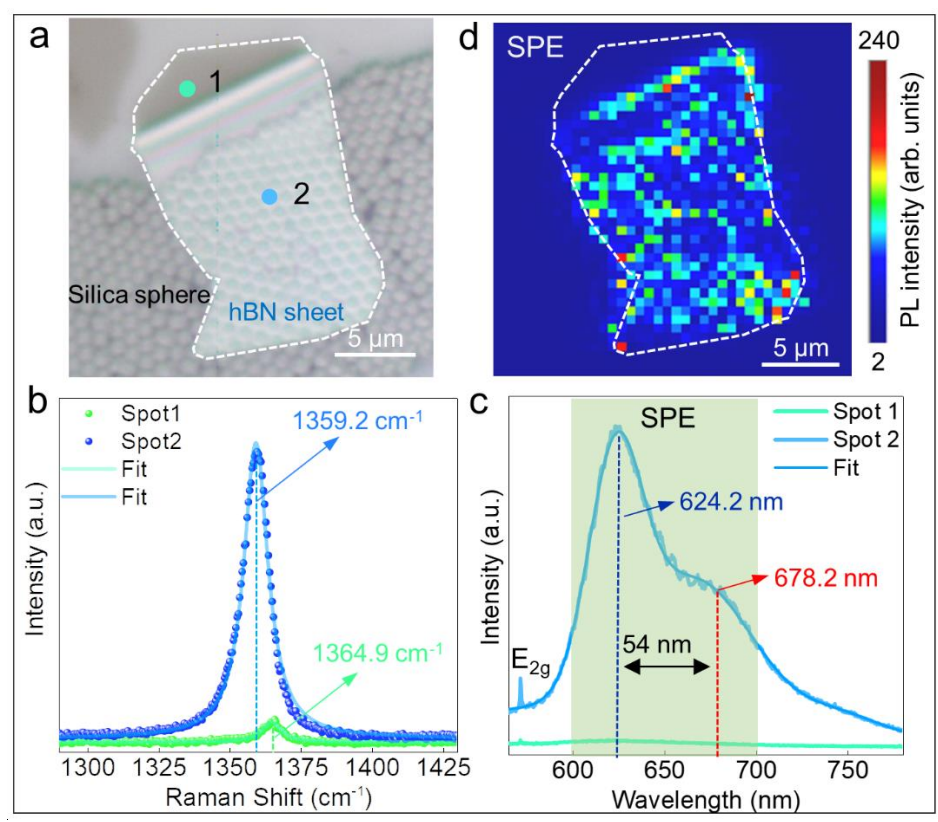

**Figure 3**

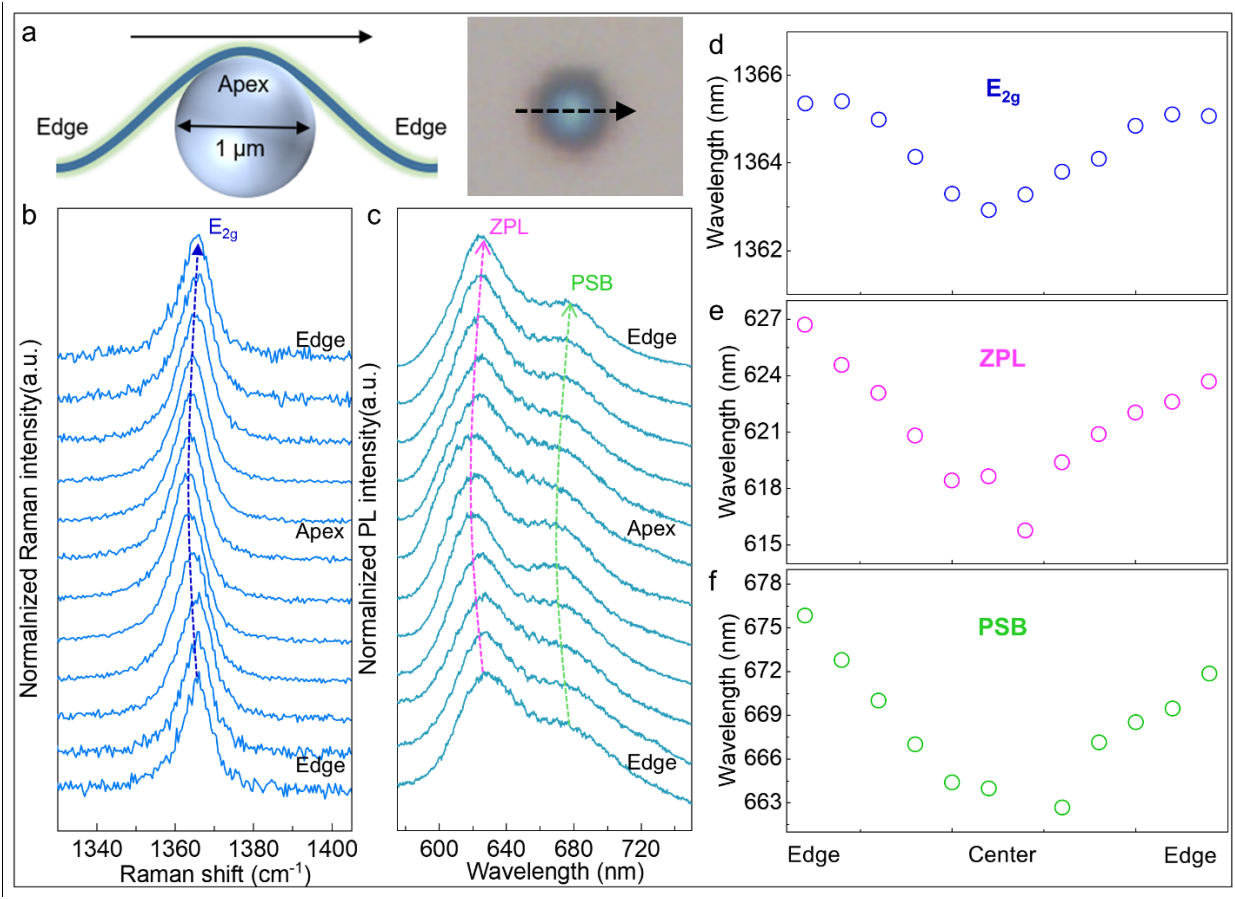

**Figure 4**

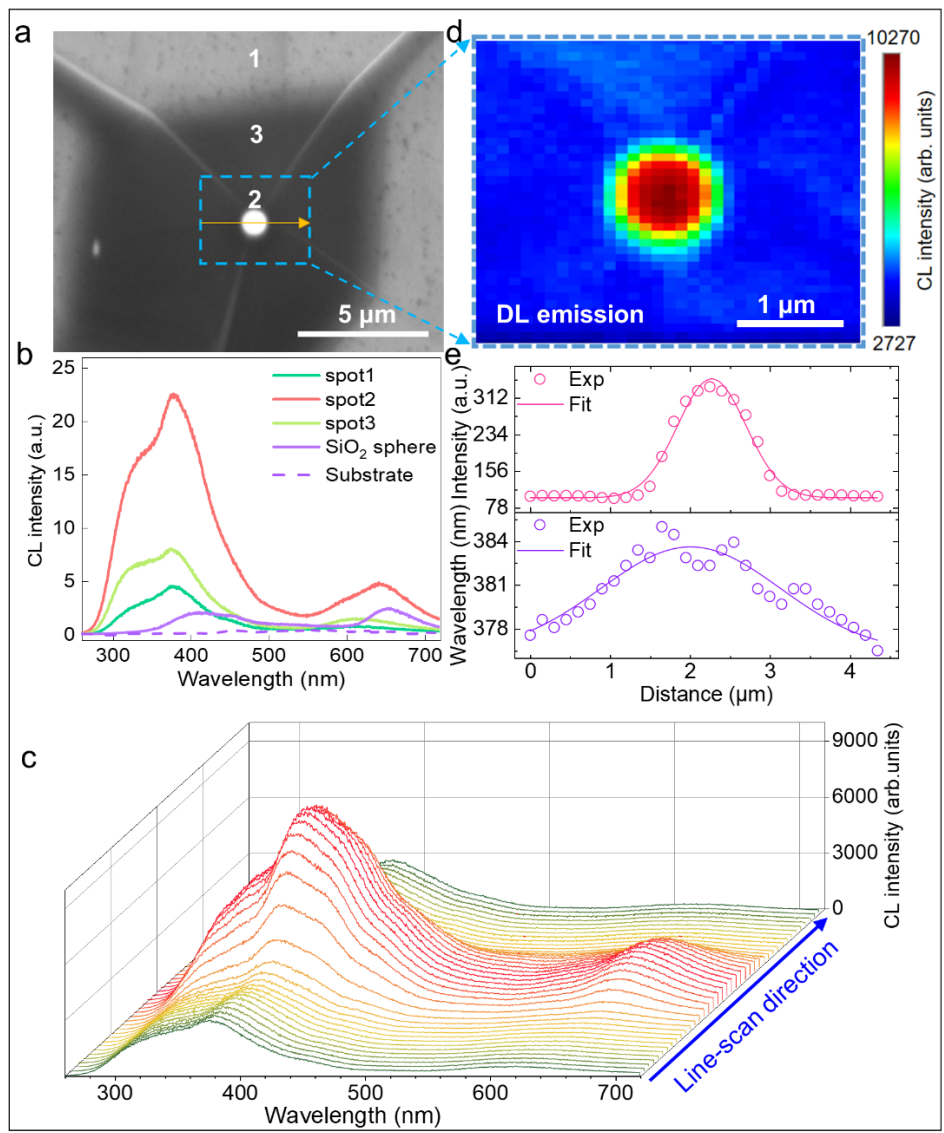

**Figure 5**

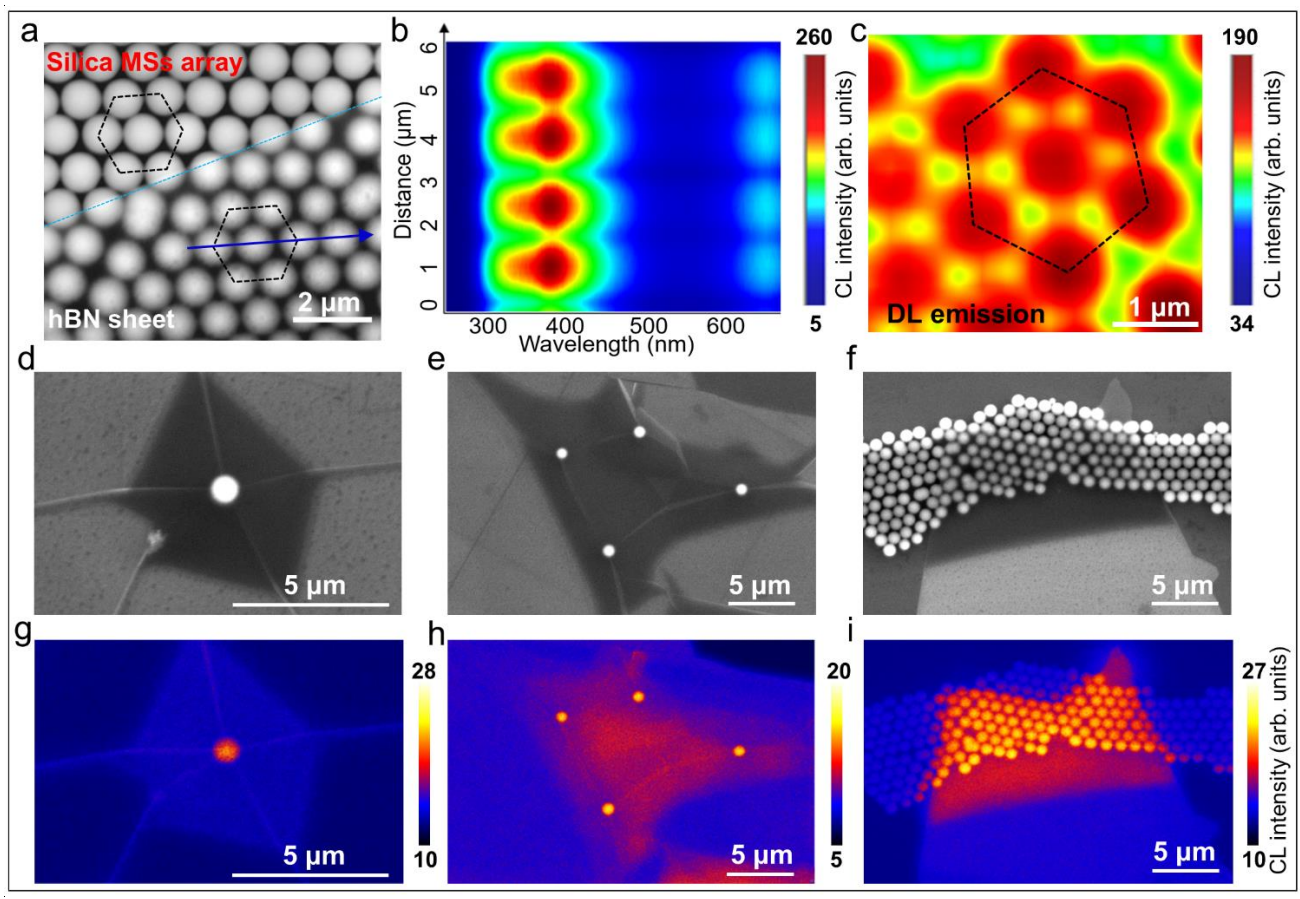